%% file: main.tex
\documentclass[10pt,conference,letterpaper]{IEEEtran} %Change this later to document class conference
\IEEEoverridecommandlockouts
\usepackage{cite}
\usepackage{amsmath,amssymb,amsfonts}
\usepackage{algorithmic}
\usepackage{graphicx}
\usepackage{textcomp}
\usepackage{xcolor}
\usepackage{xspace}

\usepackage{graphicx}
\usepackage{subcaption}
\usepackage{epstopdf}
\usepackage{float}
\usepackage{balance}
\usepackage{comment}
\usepackage{mathtools}
\usepackage{amsmath}
\usepackage{multirow}
\usepackage{xspace}
\usepackage{xcolor, color, soul}
\usepackage{url}
\usepackage{pifont}
\usepackage{wasysym}
\usepackage[noend]{algorithm2e}
\RestyleAlgo{ruled}
\SetKw{KWIn}{in}
\SetKwComment{Comment}{/*}{*/}
\DeclareMathOperator*{\argmaxA}{arg\,max}

\def\BibTeX{{\rm B\kern-.05em{\sc i\kern-.025em b}\kern-.08em
    T\kern-.1667em\lower.7ex\hbox{E}\kern-.125emX}}

\newcommand{\system}{\textit{NeTo-X}\xspace}

\newcommand{\karthik}[1]{\marginpar{\color{red}\tiny\ttfamily Karthik: #1}}

\begin{document}
\title{Scalable Network Tomography for \\ Dynamic Spectrum Access
\thanks{This work was supported in part by NSF (CNS 2208761) and Georgia Tech Research Institute.}
}
\author{\IEEEauthorblockN{Aadesh Madnaik, N. Cameron Matson, Karthikeyan Sundaresan \\}
\IEEEauthorblockA{School of Electrical and Computer Engineering, Georgia Institute of Technology, USA\\
Email: \{amadnaik3, ncmatson\}@gatech.edu, karthik@ece.gatech.edu}
}
\maketitle
% \vspace*{-10mm}
% \thispagestyle{plain}
% \pagestyle{plain}
% \pagenumbering{Arabic}
\begin{abstract}
\input{abstract.tex}
\end{abstract}
% \vspace{-3mm}
% \vspace{-2mm}
\input{introduction.tex}
% \vspace{-3mm}
\input{background.tex}
% \vspace{-2mm}
\input{system_design.tex}

% \vspace{-2mm}
\input{practice.tex}
% \vspace{-2mm}
\input{evaluation.tex}
% \vspace{-2mm}
\input{conclusions.tex}
\bibliographystyle{plain}
\bibliography{references.bib}
\end{document}

%% file: abstract.tex
Mobile networks have increased spectral efficiency through advanced multiplexing strategies that are coordinated by base stations (BS) in licensed spectrum. However, external interference on clients leads to significant performance degradation during dynamic (unlicensed) spectrum access (DSA). We introduce the notion of network tomography for DSA, whereby clients are transformed into spectrum sensors, whose joint access statistics are measured and used to account for interfering sources. Albeit promising, performing such tomography naively incurs an impractical overhead that scales exponentially with the multiplexing order of the strategies deployed -- which will only continue to grow with 5G/6G technologies. 

To this end, we propose a novel, scalable network tomography framework called \system that estimates joint client access statistics with just linear overhead, and forms a blue-print of the interference, thus enabling efficient DSA for future networks. \system's design incorporates intelligent algorithms that leverage multi-channel diversity and the spatial locality of interference impact on clients to accurately estimate the desired interference statistics from just pair-wise measurements of its clients. The merits of its framework are showcased in the context of resource management and jammer localization applications, where its performance significantly outperforms baseline approaches and closely approximates optimal performance at a scalable overhead.

%% file: introduction.tex
\section{Introduction}
\label{sec:introduction}
%\% Why is dynamic spectrum access becoming important?
% \vspace{-1mm}
\textbf{Need for DSA:} 5G and 6G mobile networks are moving to higher frequencies to access larger bandwidths in response to exponential traffic growth.  Nevertheless, because of the limited operational range of these frequencies, lower spectral bands (sub-6 GHz) have remained a highly sought after resource for mobile access. Indeed, 
%realizing several lower spectral bands that have been allocated to federal organizations have had limited use, 
the FCC has been exploring the possibility of re-farming some of these federal-owned bands, with low utilization, for mobile access. A classic, recent example is the CBRS band~\cite{cbrs_alliance}  (3.5-3.7 GHz) that is being repurposed from naval radar operation to 
%A classic, recent example is the CBRS band~\cite{cbrs_alliance} (3.5-3.7 GHz) that is being re-purposed from naval radar operation with 
% through 
various flexible models of spectrum sharing. As spectrum sharing moves towards lightly-licensed and unlicensed models, dynamic spectrum access (DSA) continues to be an important problem in our search for better use of our critical spectral resources. 

\begin{figure}
\centering
    \begin{subfigure}[b]{0.35\linewidth}
    \captionsetup{justification=centering}
        \centering
        \includegraphics[width=\textwidth]{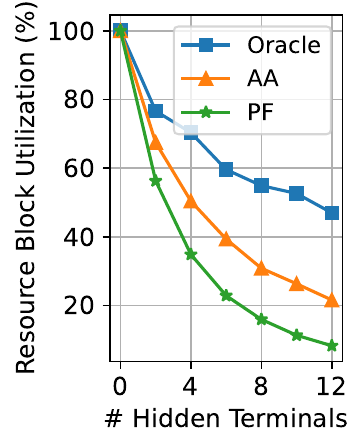}
        % \caption{Four-channel, 30 client, SISO system}
        \caption{SISO}
    \end{subfigure}
    % \hfill
    \begin{subfigure}[b]{0.63\linewidth}
    \captionsetup{justification=centering}
        \centering
        \includegraphics[width=\textwidth]{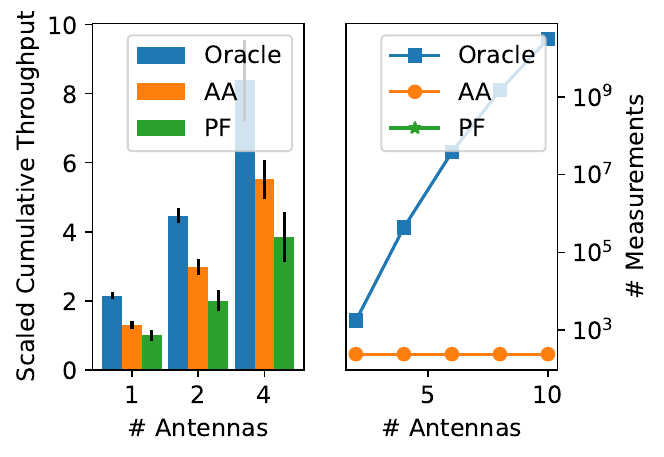}
        % \caption{Four-channel, 30 client, \newline MU-MIMO system}
        \caption{MU-MIMO}
    \end{subfigure}
    \caption{HOD prevents drop in performance with (a) increased interference, (b) larger number of antennas}
    % \vspace{-5mm}
    \label{fig:motivation}
\end{figure}

% \karthik{Fill in missing references, numbers.}
\textbf{Multiplexing gains vs. open spectrum access:} Mobile networks (LTE/5G) have evolved to accommodate several sophisticated technologies (e.g. carrier aggregation~\cite{Rapeepat-VTC'10}, multi-user MIMO~\cite{marzetta_noncooperative_2010}) that increase spectral efficiency by leveraging synchronous and licensed spectrum access. For mobile networks to be able to democratize access, the exorbitant cost of spectrum ownership needs to be tackled. 
While this can be addressed by moving towards unlicensed spectrum (e.g. LTE/5G in CBRS bands~\cite{alleven_cbrs_2020}), it sacrifices the gains these networks have worked hard to deliver.  We conduct network simulations with an LTE base station (BS), several tens of clients and a variable number of interfering sources, dubbed hidden terminals (HTs), to highlight this impact as illustrated in Fig.~\ref{fig:model}. The conventional proportional fair (PF) scheduler~\cite{kushner_convergence_2004} is deployed alongside {\em mandated} unlicensed spectrum access protocol \cite{laa_spec} at the BS. The results, summarized in Fig.~\ref{fig:motivation}, reveal: 
\begin{figure}
    \centering
    % \vspace{-5mm}
    \includegraphics[width = 0.9\linewidth]{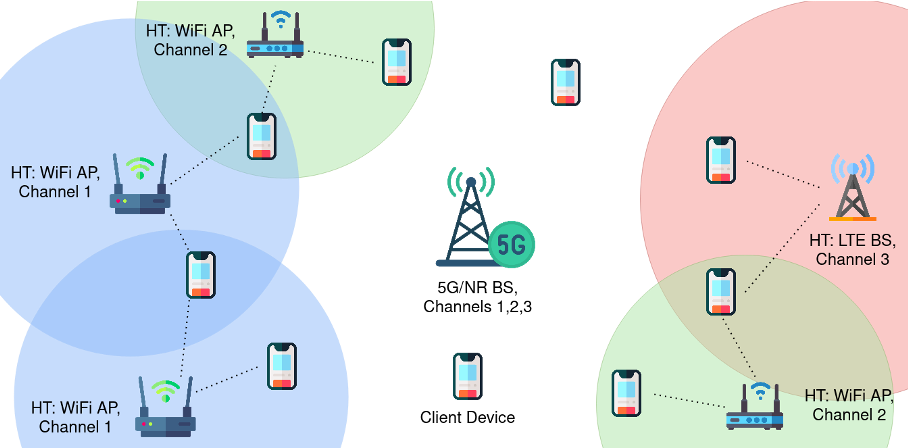}
    \caption{Illustration of hidden terminal interference}
    % \vspace{-5mm}
    \label{fig:model}
\end{figure}

(a) {\em Interference can be modelled probabilistically: } Hidden terminal interference is commonly modelled as a Poisson Point Process \cite{heath2013poisson, huang1995interference, hu2020interference}. As observed through our simulations, and verified by \cite{ruzankin2004poisson}, this interference can be closely approximated by a Bernoulli Process with discretized time intervals.

% \aadesh{Add BLU and experimental traces}
(b) {\em Interference impact is significant:} As seen in Fig.~\ref{fig:motivation}a, with an increase in the number of HTs, the  ability of the BS to deliver or receive traffic from the affected clients diminishes, thus leading to a rapid drop (as high as $80\%$) in resource utilization. This reveals existing networks' acute vulnerability to interference despite sophisticated access protocols. 

(c) {\em Limited interference information is useful, but not sufficient:} 
In contrast to PF, which is interference-agnostic, we consider an alternate {\em access-aware} (AA) scheme. It incorporates some knowledge of the interference---namely the probability that individual clients can access the channel--- in the scheduling decisions. 
%We consider both SISO and MU-MIMO performance in Figs.~\ref{fig:motivation}(a) and (b) respectively. 
The AA scheme performs 50-100\% better than PF, which is promising. However, as more antennas $M$ are introduced in our 5G systems, the throughput depends on the specific set of users that are jointly scheduled on the same uplink resource. Hence,
%in addition to the access probabilities of `individual' clients considered in AA, 
% \aadesh{One more sentence here to convey HOD importance}
we also consider an Oracle that has access to all higher-order `joint' access distributions (HODs) of all combinations of clients. Such an ideal scheduler benefits from interference-aware `group' selection that is critical for MU-MIMO, and made possible by HODs. Fig.~\ref{fig:motivation} reveals that while AA improves performance, there is still a wide gap to the theoretical performance that is possible with HODs. 

% \karthik{Why does PF grow linearly with antennas even in the presence of interference in Fig.1b.}
% (iii) {\em HOD are essential to retain multiplexing gain in MU-MIMO systems:}The MU-MIMO version of the PF and AA schedulers do not consider the ability of the clients selected to `simultaneously' use the resources in the presence of interference. Hence, as the multiplexing order ($M$) increases in Fig.~\ref{fig:motivation}(b), this has an even greater impact(compared to the SISO system), severely limiting the multiplexing gain that could be achieved, as is observed with respect to Oracle's performance -- the latter benefiting from interference-aware `group' selection that is critical for MU-MIMO, and made possible by HOD.

% (iv) {\em HODs useful even in SISO systems:}
% In a SISO system, why might we need access to the HOD? Consider that in AA, once a client is scheduled, it is possible for the allocated resources to be completely wasted if there is interference at the client. Having access to HODs, allows the Oracle to speculatively \textit{over}-schedule multiple clients (on uplink) on the \textit{same}, single antenna resource.  By using the HOD, it intelligently selects clients so as to maximize the expected resource utilization over time while minimizing the probability of collisions. This results in a much more graceful degradation in the Oracle's performance with increasing interference, compared to AA in the SISO system in Fig.~\ref{fig:motivation}(a). We discuss this approach in Sec.~\ref{sec:sched_sp}. 

% \aadesh{Try to be more concise here}
\textbf{Network Tomography for DSA:} An important step in addressing such unknown interference is to first estimate it accurately. However, estimating interference is, in fact, a (receiver) \textit{location}-dependent one. Thus, even a sophisticated spectrum scanning solution located at the BS cannot obtain a comprehensive view of the interference environment. This has led to the rise of network tomography for DSA~\cite{blu}, whereby the client devices' existing communication interfaces double-up as virtual spectrum sensors. Through intelligent scheduling of the clients and measuring the corresponding outcomes, the BS can decipher the impact of interference on the clients, accounting for both spectrum and location dependence. 

%While spectrum sensing has understandably become a key component of DSA research~\cite{ahmad_5g_2020,Chen2015offload, 5gnr_unlicensed, qualcomm-lte-u, qualcomm_laa}, estimating interference is not just a spectrum-dependent process; it is equally a (receiver) \textit{location}-dependent one. This makes it extremely challenging for even a sophisticated spectrum scanning solution at the BS to obtain a comprehensive view of the interference its clients experience. This has led to the rise of network tomography for DSA~\cite{blu}, whereby the client devices' existing communication interfaces double-up as virtual spectrum sensors. Through intelligent scheduling of the clients and measuring the corresponding outcomes, the BS can decipher the impact of interference on the clients, accounting for both spectrum and location dependence. 

\textbf{Challenge of Scalable Tomography:} 
While HODs enables a network to operate in a smarter, more efficient way as seen in Fig.~\ref{fig:motivation}, the challenge lies in acquiring them in the first place. However, a naive approach would reveal only the first-order \emph{marginal} probabilities. 
%We have seen (Fig.~\ref{fig:motivation}) that we require the HOD to operate the network efficiently.  
The HOD answer questions such as ``What is the probability that clients $i$ and $j$ are interfered with but clients $k$ and $l$ are not?"
We could continue to measure the HOD directly by scheduling each possible combination of clients, but in an $M$-antenna, $C$-channel MU-MIMO system with carrier aggregation~\cite{CA}, this results in an exponential measurement overhead $O(C \cdot 2^M)$, as seen in Fig.~\ref{fig:motivation}b, which wastes spectral resources. 
%\aadesh{Condense} Note that, if the system employs multiple channels ($C$), as is the common case today with carrier aggregation~\cite{CA}, then the overhead scales as $O(C\cdot 2^M)$. This overhead makes a true Oracle intractable in practice.
%How would one go about measuring these probabilities?  The number of measurements required is obviously a function of the number of users served by the network, but also the number of antennas and the number of channels.  As each of these parameters increases, so to does the number of measurements required to accurately estimate the desired probability.
%Fig.~\ref{fig:motivation}(b) demonstrates this challenge by plotting the number of measurements required to estimate the desired HOD for an increasing number of antennas.  
%While the AA scheme does not rely on HODs, the number of measurements needed by the Oracle to estimate and benefit from HOD increases exponentially, making it impractical.

%To address this fundamental challenge, our proposed a network tomography approach for DSA, \system, allows us to leverage the benefits of HODs, not just for scalable performance in higher-order antenna and channel systems, but also in other interference-analysis applications (e.g. jammer localization)---all at a practical overhead that scales linearly with the number of channels employed and the clients ($O(C \cdot N + C\cdot {K \choose 2})$). 
 
% \aadesh{Double blind so don't state earlier work}

In this work, we aim to address a fundamental challenge: {\em Can we continue to realize mobile network's multiplexing features, while maintaining performance comparable to licensed spectrum access, even in unlicensed spectrum with unknown interference with low overheads?} 

\textbf{\system:} To this end, we propose a novel, \underline{scalable} \underline{ne}twork \underline{to}mography framework called \system that estimates HODs more accurately, at just a linear overhead, and generates a blue-print of the interferers, to enable efficient DSA for 5G/6G networks. \system helps address the scalability challenge in accurately estimating HODs across multiple channels, antennas, and a large number of clients. The HOD serves as a key building block for multiple applications like (i) resource management---enabling mobile networks to retain the benefits of their multiplexing techniques, and (ii) security---allowing defense and private networks to localize potential interfering/jamming sources.

\textbf{Design:} \system's design incorporates three key elements:
% \karthik{choose representatives for what? clarify a little more}

(a) {\em Interference-Aware Clustering:} \system leverages the spatial consistency and channel diversity of the interference experienced by clients to cluster clients and choose representatives. A single representative client from a cluster is sufficient for measurements, which helps to significantly reduce the measurement overhead. % Clustering significantly reduces the overhead dependence  from $N$ clients to $K \ll N$ clusters, independent of the number of clients, channels or antennas, where a single representative client from a cluster is sufficient for the measurements.
% \aadesh{Add small detail about K}

(b) {\em Generating HODs from Pairwise Measurements:} Leveraging recent results in latent variable decomposition, \system proposes an algorithm to estimate the HODs of arbitrary groups of clients, from just pairwise measurements of the clusters' representatives. %, resulting in an aggregate overhead of $O\big(C \binom{K}{2}\big)$ that is independent of both the number of clients $N$ and antennas $M$, and scales linearly with  $C$ channels.
%\karthik{We are still using both C,K and C,N in introduction. Make them consistent. Also, check in Sec 4}
%\aadesh{This one is correct. It talks only about the HOD estimation from pairwise}

(c) {\em Blue-Printing Interference:}  \system proposes a novel algorithm that accurately estimates the number of interfering sources and their corresponding dependencies on clients based on the latent variable decomposition of the HODs. %, along with the localized nature of interference impact.
% \karthik{Update for correctness}

\textbf{Applications:} \system showcases two key applications.

(a) {\em Resource Management:}  The HODs are incorporated into mobile network's scheduling framework to enable interference-aware multi-user access. %, maximizing the multiplexing gain even in the presence of interference.
% whereby higher dimensional MU-MIMO transmissions across multiple channels are effectively leveraged even in the presence of external interference through intelligent, interference-aware client grouping.   
\system is implemented in NS3~\cite{ns3} and supplemented with numerical evaluations to understand \system's ability to deliver gains in multiplexing (higher order MU-MIMO and carrier aggregation) schemes despite the presence of external interference. Our evaluations highlight that \system helps deliver $> 2\text{x}$ throughput gain over existing schedulers in high interference regimes for SISO, while maintaining a gain of $2.5\text{x}$ even for low-moderate interference regimes for MU-MIMO systems.

(b) {\em Interferer/Jammer Localization: } Leveraging the blue-printed interference, \system is able to identify a small set of key clients, whose locations in turn allow us to maximally narrow down the location of the interfering sources to a relatively small area of under $200~m^2$. Further, \system localizes hidden terminals to a 75-th percentile error of under 10m - a feature not possible by existing DSA schemes.

Our contributions in this work are as follows. 

\noindent {$\bullet$} We propose a scalable network tomography framework \system that enables efficient multi-user, multi-channel access for 5G/6G systems in unlicensed spectrum at low overhead. 

\noindent {$\bullet$} \system designs a novel, interference-aware channel allocation and resource management mechanism that helps deliver the benefits of carrier aggregation and higher-order MU-MIMO even in the presence of interference. 

\noindent {$\bullet$} \system proposes a novel mechanism that further leverages the hidden dependencies between interfering sources and clients, to localize the interfering sources from a carefully identified, small set of clients and their positions. 
%our work makes the following key contributions.
%We propose a novel network tomography solution that  
%\% what is the contribution of this work over prior art? What are the key contributions?

%% file: background.tex
\section{Background and Related Work}
% \vspace{-2mm}
\textbf{5G Access Overview:} Cellular networks (Long Term Evolution, LTE in 4G and upcoming New Radio, NR in 5G) are synchronous, scheduled access systems designed for operation in the licensed spectrum.
The base station (BS) is responsible for scheduling both the downlink (DL) and uplink (UL) clients in its sub-frames, which  consists of two-dimensional resource elements spanning both time (symbols) and frequency (sub-carriers), called resource blocks (RBs).
LTE/5G-NR employs OFDMA (orthogonal frequency division multiple access), whereby multiple clients are scheduled in each sub-frame on {\em different} RBs - in the case of multi-user and massive MIMO that are central to 5G systems, multiple clients are scheduled on the {\em same} RB, but spatially separated by multiple antennas. The schedule for both DL and UL transmissions is conveyed to the clients through the control part of the DL sub-frames.

 % \karthik{Need more recent relevant references.}
\textbf{5G in Un-licensed Spectrum:} Unlike traditional cellular systems that operate in an always-on mode in licensed spectrum, operating in unlicensed spectrum {\em requires} 5G to adopt asynchronous access principles of clear-channel assessment (CCA) through energy sensing, called Listen-Before-Talk~\cite{laa_spec}
and back-off for co-existence with other technologies (e.g. WiFi or other cellular networks). 
 %the incumbents. 
The prevalent solution today is license-assisted access (LTE-U~\cite{lteu_forum} and LTE-LAA/eLAA~\cite{3gpp_laa,qualcomm_laa,laa_spec}), along with the numerous works that have studied the LTE-WiFi coexistence problem~\cite{5gnr_unlicensed, naik2022_wifi_5gnru, Rupasinghe2015,Mukherjee2015,sagari2015,Chen2015offload,Yun2015}, where unlicensed carriers are aggregated with existing licensed carriers. By relying on licensed carriers as anchors, the impact of asynchronous interference is confined to unlicensed carriers, thus limiting the overall impact.
Similarly, the use of spectrum access servers in CBRS~\cite{deepradar} enables coarse timescale coordination for exclusive use of the spectrum.
 % \karthik{People might confuse our work with SAS based spectrum sharing approaches. I added a sentence/ref for that.}
In contrast, operating 5G entirely in unlicensed spectrum (e.g. MulteFire~\cite{multefire_spec} in 3.5 GHz CBRS bands in GAA mode) that is shared across multiple entities, exposes the vulnerability of its capacity-enhancing multi-user schemes to asynchronous interference (e.g. WiFi APs interfering with clients but not sensed by the BS), highlighting its lack of readiness for practical deployments~\cite{blu}. 
% \aadesh{Edit BLU, 33} 
While \cite{blu} has showed promise for how 4G LTE systems can cope with such asynchronous interference, it left open several fundamental algorithmic and system level challenges that must be solved for a practical 5G solution, especially scaling to larger multiplexing gains (from larger bandwidths and antenna arrays in 5G) during real-time operations. This, in turn, forms one of the objectives of this work. 

\textbf{Network Tomography for DSA:} Spectrum sensing has understandably become a key component of DSA research~\cite{ahmad_5g_2020,Chen2015offload, 5gnr_unlicensed, qualcomm-lte-u, qualcomm_laa}, but estimating interference is not just a spectrum-dependent process. Meanwhile, network tomography has been popular in wired networks~\cite{yaseen_synchronized_2018, ben_basat_pint_2020, brugere_network_2018, bu_network_2002} where it refers to leveraging measurements between end-hosts of a network so as to reveal important properties of the network itself. \cite{blu} introduced a similar notion for DSA, whereby they transformed the clients of our cellular network into spectrum sensors, whose measurements in accessing the spectrum are leveraged to reveal interesting properties on the interference they face. This is made possible by OFDMA, where a BS can orchestrate measurements by scheduling clients either in isolation or simultaneously with others and estimating the resulting access distributions.    

%\subsection{LTE/NR overview}

%\karthik{This should be brief to just capture synchronous DL/UL transmissions with a figure. Move scheduling background to later. Limit background+related work to under 1 page for section 2.}

%\subsection{LTE/NR in un-licensed spectrum}
%\% cover related work, while also mentioning and illustrating how hidden terminal interference impacts cellular networks that are not designed to handle it. 

%\subsection{Network Tomography}
%\% network tomography as a concept (very brief)

%% file: system_design.tex
\section{System Design}
\label{sec:system_design}
\begin{figure}
    \centering
    % \vspace{-2mm}
    \includegraphics[width = \linewidth]
    {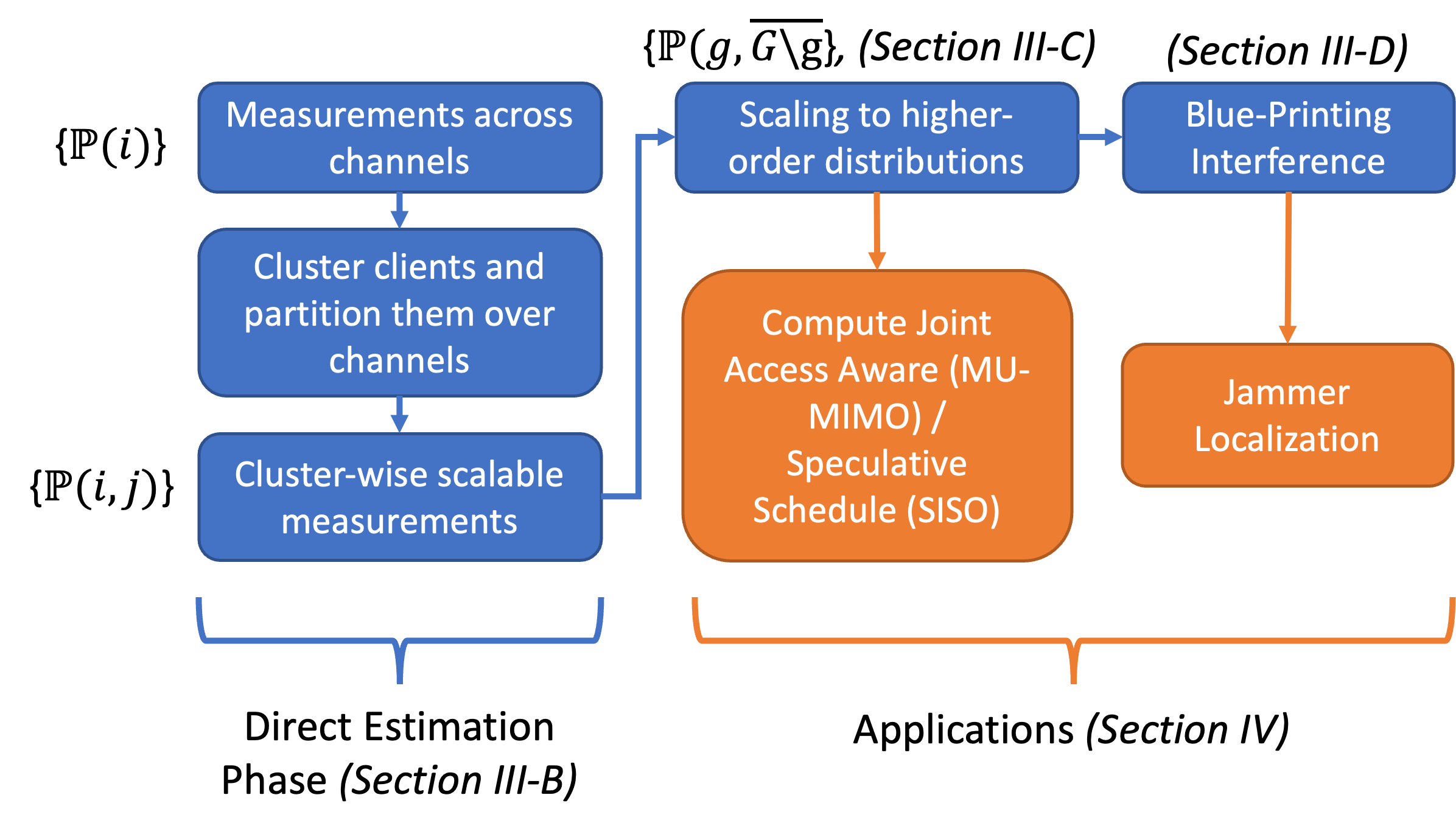}
    % \vspace{-2mm}
    \caption{System Design and Information Flow}
    % \vspace{-6mm}
    \label{fig:flow}
\end{figure}
% \vspace{-3mm}
\subsection{Overview}
\label{sec:system_design_overview}
% \vspace{-2mm}
As stated in Section~\ref{sec:introduction}, having complete information about the joint-access distribution (HODs) is beneficial for efficient resource management as well as localizing interfering sources. To this end, the goal of \system is to accurately estimate HODs with as few measurements as possible and without interrupting the normal operation of the network. %\aadesh{Restate the problem mentioned in the introduction. Talk about contributions and how it ties in here}
\system does this by treating the clients as a distributed set of sensors to learn information about the interference environment and intelligently scale those measurements to estimate the HODs.

The system consists of several steps summarized in Fig.~\ref{fig:flow}.
% In each phase the goal is to maximize the amount of information gained from  as few measurements as possible.
% without impacting the normal operation of the network.
\system's operation is divided into two phases: a \textit{direct measurement} phase and a \textit{scaling} phase.
The measurement phase consists of three steps: measuring first-order marginals per client, channel-wise clustering of clients, and measuring second-order marginals per cluster.
First, the first-order marginals are measured during clients' ongoing data transmissions by observing the interference they experience on each channel (Section~\ref{sec:channelsampling}). 
Next, to reduce the amount of measurements required in estimating second-order marginals, \system uses a clustering algorithm to group clients with similar interference patterns such that interference dependencies can be captured at a cluster- (instead of client-) level without losing information  (Section~\ref{sec:clustering}).
Finally, 
second-order marginals are measured by scheduling pairs of clients from unique clusters (Section~\ref{sec:pairwise}).
% we select a representative from each cluster on each of the channels ($C$) to estimate the second-order (pair-wise) marginal probabilities (Section~\ref{sec:pairwise}) between clusters in the respective channels. 
% This results in an $O(C\cdot K^2)$ overhead, where $K$ is the number of clusters and is independent of number of clients or antennas.

In the second phase (Section~\ref{sec:scale}), we use latent variable decomposition to scale these estimates up to the HOD.
% in a provably optimal manner\hl{Check}.
% The method relies on uncovering dependencies between the measurements and a set of hidden latent variables. 
Once the HODs have been estimated, we can use these for client scheduling as discussed in Section~\ref{sec:introduction}.  An additional step (Section~\ref{sec:blueprinting}), which we call interference blue-printing, uses the HOD 
% to reveal more information about the interference itself, allowing us 
to uniquely identify the sources of interference. This information, along with knowledge of a few client locations, can be used to localize the interferers themselves. 

\textbf{Notation: } 
% All vectors are represented using bold and lowercase lettering (like $\textbf{a}_i$). 
% All matrices are represented with bold and capital lettering (like $\textbf{A}_n$).
% All tensors are represented with bold, capital lettering with a bar under (like \textbf{\underline{X}}).
Vectors, matrices and tensors are represented as $\mathbf{a}$, $\mathbf{A}$, $\mathbf{\underline{X}}$, respectively.
The $i$-th element of a vector is given by $\mathbf{a}(i)$.
Similarly, $f$-th column and $i$-th row of a matrix $\mathbf{A}$ are notated $\mathbf{A}(:\,,f)$ and $\mathbf{A}(i, :)$, respectively.  

% \vspace{-1mm}
\subsection{Direct Estimation Phase}
    % \vspace{-1mm}
    \subsubsection{Channel Sampling}
    \label{sec:channelsampling}
    % single order marginal
    The first phase of \system consists of estimating the channel access probability for each client individually across all of the channels.
    The probability that a client $i\in[1,\ldots,N]$ is able to access channel $c \in [1,\ldots,C]$ is described by a Bernouli random variable with parameter $a_i^c$.
    We estimate these individual channel accesses probabilities empirically by counting the number of times a scheduled client is able to access the channel.
    \begin{align*}
    \small
        \mathbb{P}(\text{client $i$ can access channel $c$}) = \frac{\text{\# times $i$ accesses $c$}}{\text{\# times $i$ scheduled on $c$}} = a_i^c %\\
        %&= a_i^c
        \label{eq:est_1order}
    \end{align*}
    By stacking each of the channel access probabilities for a given client, we create a channel access vector $\mathbf{a_i} = [a_i^1, a_i^2, \ldots, a_i^C]$.
    This process requires scheduling each of the $N$ clients across each of the $C$ channels to take measurements, i.e. total overhead of $O(CN)$. Measurements are collected through uplink grants, and the clients send uplink packets as part of the normal operation of the network.
    
    \subsubsection{Client Clustering}
    \label{sec:clustering}
    The direct phase continues by measuring pair-wise access probabilities.  Doing so naively would require us to schedule every pair of clients (jointly) on each of the channels, resulting in an overhead of $O(C \cdot \binom{N}{2})$ measurements.
    To reduce this number, we leverage the following observation: 
    if two clients are physically located close enough such that they are affected by the same set of HTs, they will likely have similar first-degree marginal channel access probabilities over all of the channels. 
\begin{figure}
    \centering
    \includegraphics[width = \linewidth]{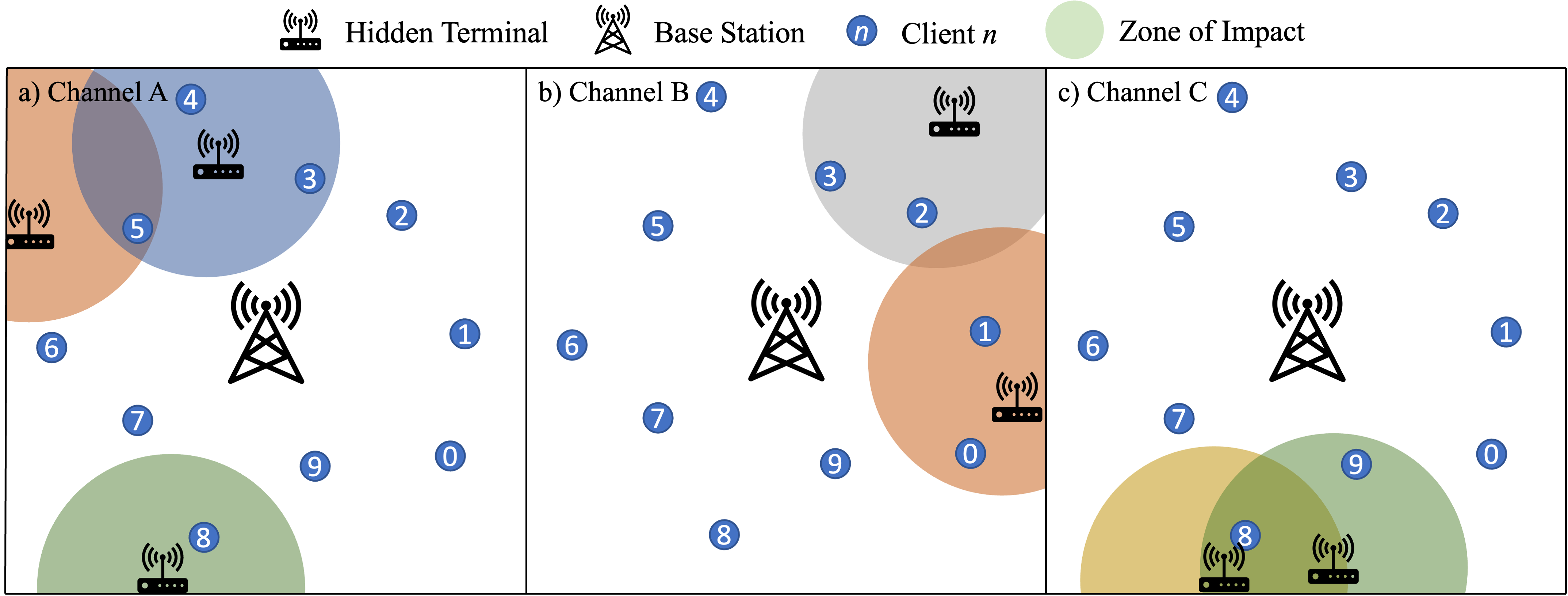}
    \caption{Impact of hidden terminals across three channels.}
    \label{fig:HT-channel}
    % \vspace{-8mm}
\end{figure}
To illustrate this, consider Fig.~\ref{fig:HT-channel}.  In this topology there are ten clients, seven HTs, and three channels. 
    % Through this figure we illustrate different client-HT and HT-HT interactions. 
    For example, in channel $A$, client $5$ is impacted independently by two HTs. 
    However, in channel $C$, the two HTs are close enough such that their transmissions can be sensed, through some form of carrier sensing like CSMA, by one another. So, when the yellow HT is transmitting, the green HT cannot use the channel (which frees client 9 to transmit). In this scenario, client 8 is impacted by a \textit{dependent} set of HTs.
    Now, consider clients 0 and 1. After the first phase of measurements, we would observe that they have similar channel access probabilities on all channels---no interference on channels $A$ and $C$, and same probability of interference on channel B, i.e. %The result is that 
    $\mathbf{a}_0 \approx \mathbf{a}_1$.  
    %It is important to note that even though client 9 is approximately the same distance away from client 0 as client 1, it observes a different interference pattern, demonstrating that physical proximity of clients does not guarantee they will experience the same interference. 
    %Hence, the larger the channel access vector 
    
    We group the clients into clusters via K-means using the squared $L2$ norm of the difference between their channel access vectors, $\mathbf{a}$, to measure the distance. Note that clustering based on identifiers like RSSI has previously been used to spatially locate clients \cite{pinto2021rssi}.
    After clustering it is sufficient to perform pair-wise measurements with a single representative client from each cluster. Intuitively, this is because the clusters group clients that experience similar interference; therefore, the joint distributions of the representative clients across the clusters will be the same as those between the remaining, non-representative clients from those same clusters.  
    The clustering phase reduces the size of measurement set from $N$ clients to $K$ clusters which does not depend directly on $N$. 
    We note that $K$ is a design parameter of \system. An approximation is sufficient, since $K$ should simply capture number of the contention (interference) regions in the network. 
    We investigate the impact of incorrect clustering as well as the choice of $K$ on the accuracy of the eventual HOD and the measurement overhead reduction in Section~\ref{sec:cluster_analysis}. 
    % \aadesh{Here, also talk about what happens when the clustering is incorrect}
    
    % 2 clients that are close in this space would likely not benefit of being scheduled at the same time because there is a high probability that both or neither would be able to access the channel
    
    \subsubsection{Pairwise Probability Estimation}
    \label{sec:pairwise}
    The final step of the direct estimation phase is to measure the pair-wise channel access which can be computed empirically as before with the single client access.
    The BS simultaneously schedules a representative client from distinct clusters $i$ and $j$ and counts the number of opportunities one, both, or neither are able to access the channel.  This incurs $O\big(C\binom{K}{2}\big)$ measurements, after which we have an estimate of all (cluster) pair-wise channel access probabilities, $\mathbb{P}(i, j)$.
% \vspace{-2mm}
\subsection{Scaling to High Order Distributions}
\label{sec:scale}
% \vspace{-1mm}
Clustering helps to keep the measurement set manageable for estimating the pair-wise densities, but the number of measurements required to estimate HODs directly---even between clusters---quickly becomes unmanageable.
% Table~\ref{tab:overhead}  summarizes the number of measurements per channel\footnote{The space requirements to store the PMF are proportional to the number of measurements.} required to estimate various degree distributions.
In order to scale the pair-wise marginals up to higher order marginals of arbitrary size (including full joint distribution) we leverage latent variable decomposition techniques from~\cite{pairwisetensor,threewaytensor}.

    \subsubsection{Joint Probability Representation}
    \label{sec:hod_primer}
    Consider a set of discrete, finite-alphabet random variables $Z_1, Z_2, \ldots, Z_N$. We use $\mathbb{P} \left( z_1, z_2, \cdots, z_N \right)$ as a shorthand notation to represent $\mathbb{P}(Z_1 = z_1, Z_2 = z_2, \ldots, Z_N = z_N)$.
    %where $\{z_n^{(1)}, z_n^{(2)}, \ldots, z_n^{(|Z_n|)} \}$ is the alphabet set for $Z_n$ and $z_n$ is an index into that alphabet. 
    Consider the scenario in which we do not have access to the joint probability distribution for $\{Z_n\}_{n=1}^N$.  Rather, we only have access to one- and two-degree marginals, $\mathbb{P}(z_i, z_j)$ over all pairs $\{i,j\}$. 
    %{\color{red} Ibrahim et al.}~\cite{} asks the question 
    This leads to an important question:
    
    \textit{Can we recover the joint probability distribution given we have access to solely low-dimensional marginals without any structural assumptions on the random variables?} 

    We can use the fact that any joint probability mass function (PMF) admits a Naive Bayes' model representation. This implies that any joint PMF can be generated from a latent variable model with just one hidden variable ($H$) given a sufficiently rich alphabet ($F$, \cite{pairwisetensor} Theorem 1). Thus, the joint PMF of $\{Z_n\}_{n=1}^N$ can always be decomposed as, 
    % \vspace{-2mm}
    \begin{equation}
    	\mathbb{P}(z_1, \ldots, z_N) = \small \sum_{f=1}^F \mathbb{P}(H=f) \prod_{n=1}^N \mathbb{P}(Z_n = z_n | H=f).
     % \vspace{-2mm}
     \label{eq:PMF}
    \end{equation}
    We exploit this property to express degree-2 marginals as
    % \vspace{-2mm}
    \begin{equation}
        \label{eq:pair_decomp}
        \mathbb{P}_{ab}(z_a, z_b) = \sum_{f=1}^F \boldsymbol\lambda(f) \mathbb{P}(Z_a = z_a|f) \mathbb{P}(Z_b = z_b|f).
        % \vspace{-2mm}
    \end{equation}
    % and higher-order marginals, e.g. of four variables, as:
    % \vspace{-3mm}
    % \begin{equation}
    %     \mathbb{P}_{abcd}(z_a, z_b, z_c, z_d) = \sum_{f=1}^F \boldsymbol\lambda(f) \prod_{i \in \{a,b,c,d\}} \mathbb{P}(Z_i = z_i | f)
    %     \vspace{-1mm}
    %     \label{eq:4way}
    % \end{equation}
    The correctness of Eq.~\ref{eq:PMF} is related to canonical polyadic decomposition (CPD). Given an N-way tensor $\underline{\textbf{X}} \in \mathbb{R}^{I_1 \times \cdots \times I_N}$ with CP rank $F$, it admits a decomposition of the form: 
    % \vspace*{-2mm}
    \begin{equation}
       \underline{\textbf{X}} = \sum_{f=1}^{F} \boldsymbol\lambda(f) \textbf{A}_1(:, f) \otimes \textbf{A}_2(:, f) \otimes \cdots \otimes \textbf{A}_N(:, f)
       % \vspace{-2mm}
       \label{eq:CPD}
    \end{equation}
    where $\otimes$ is the tensor outer product operator, $\textbf{A}_n \in \mathbb{R}^{I_n \times F}$ is called the mode-\textit{n} latent factor, and $\|\boldsymbol\lambda\|_0 = F$ is used to normalize the columns of $\textbf{A}_n$. Mapping Eq.~ \ref{eq:CPD} onto \ref{eq:PMF}, as explored in \cite{pairwisetensor,threewaytensor}, indicates that the random variables $Z_1,\ldots, Z_N$ are conditionally independent given a latent variable $H$. 

    \subsubsection{From Pairwise to HOD}
    \label{sec:pair_to_hod}
    In the case of channel access probability, each of the random variables is binary ($z_n \in \{0,1\}, \forall n$), so it suffices to write the degree-2 marginal for a pair of clients $i$ and $j$ as $\mathbb{P}_{ij}(\cdot,\cdot)$.
    Using Eq.~\ref{eq:pair_decomp} we represent the joint access probability of clients $i$ and $j$ using the conditionals on the latent variable $H$. As shorthand notation, we use $p_i^f$ for $\mathbb{P}(Z_i = 1|H=f)$. There are $N \times F$ independent parameters for a given set of clients and latent variable $H$. We stack them to create the matrix $\textbf{P}$. % such that $p_i^f$ is the element in the $i$th row and $f$th column of $\textbf{P}$.  
    The probability that clients $i$ and $j$ will be able to access the channel jointly can now be written as 
    % \vspace{-2mm}
    \begin{equation}
    \mathbb{P}_{ij}(1, 1) = \sum_{f=1}^F \boldsymbol\lambda(f) p_i^f p_j^f.
    % \vspace{-2mm}
    \end{equation}
    Similarly, the probability that client $i$ can access the channel, while $j$ cannot is 
    % \vspace{-3mm}
    \begin{equation}
    \mathbb{P}_{ij}(1, 0) = \sum_{f=1}^F \boldsymbol\lambda(f) p_i^f (1-p_j^f).
    % \vspace{-2mm}
    \end{equation}
    We observe that the probability measure $\mathbb{P}_{ij}$ is parameterized by $p_i^f, p_j^f$, and the vector $\boldsymbol\lambda$. Next, we minimize the KL-divergence between the measured pair-wise marginals $\hat{\mathbb{P}}(i,j)$ and the model $\mathbb{P}(i,j)$ for all pairs $\{i, j\}$. The loss function is 
    % \vspace{-2mm}
    \begin{equation}
    \mathcal{L} = \sum_{\text{all pairs } \{i,j\}} \mathbb{D}_{KL} (\hat{\mathbb{P}}_{ij},  \mathbb{P}_{ij}).
    % \vspace{-1mm}
    \end{equation}
    Under the naive Bayes' model assumption, the problem is a convex optimization problem with one hyper-parameter, the alphabet size $F$ of the hidden variable $H$.
    Thus, we can use gradient descent to find $\boldsymbol{\lambda}^*$ and $\mathbf{P}^*$.

    The power of this technique is that it allows us to write any marginal distribution of arbitrary degree using Eq.~\ref{eq:PMF}.  Consider we have a group of clients $G$ and we want to know the probability that a subset of that group $g \subset G$ will be able to access the channel while the remainder of $G$ will not.  This can be scalably found by:
    % \vspace*{-2mm}
    \begin{equation}
        \mathbb{P}(g, \overline{G\backslash g}) = \sum_{f=1}^F \boldsymbol\lambda(f)\prod_{i\in g} p_i^f \prod_{j \in G\backslash g} (1-p_j^f) 
        % \vspace{-2mm}
    \end{equation}
    
    \subsubsection{High Order Distribution Accuracy Analysis}
    \label{sec:hod_accuracy}
    This technique attempts to strike a balance between measurement overhead and the accuracy of the estimated HOD.
    It would simply not be possible to collect enough measurements to estimate the full distribution, nor would we be able to store the $2^N$ values for any moderate size value of $N$.
    On the other hand, if we start with a higher low-order marginal (e.g. three-way marginals instead of two) we could in principal increase the accuracy of the estimated HOD~\cite{threewaytensor}.
    In practice, 
    we find that using the pair-wise marginals as inputs is sufficient to achieve accurate HOD.
    Crucially, the modest gains provided by a higher order direct estimation do not outweigh the measurement overhead cost associated with them.

\begin{figure}
        \centering
        \includegraphics[width = 0.6\linewidth]{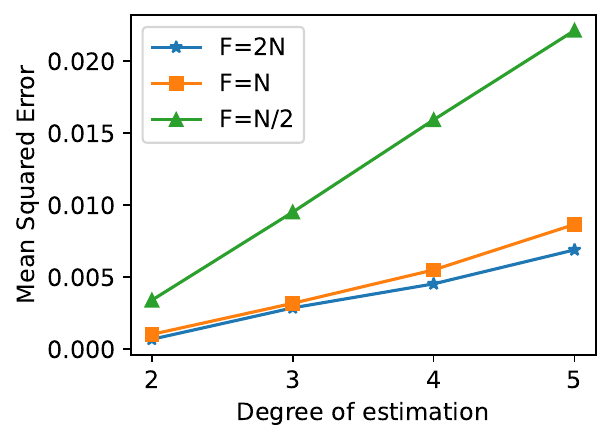}
        \caption{Mean Squared Error between true and estimated HOD from pair-wise marginal inputs vs. latent variable size.}
        % \vspace{-5mm}
        \label{fig:mse}
    \end{figure}

    Fig.~\ref{fig:mse} shows the mean squared error between the true and estimated HOD, for  $N=20$ over 50 randomized seed values, with various latent variable sizes $F$.  The estimated HODs are calculated using the measured pair-wise marginals.
    On increasing the alphabet size of the latent/hidden variable, we see that the higher-order marginals are estimated with lower error. However, the error does not reduce substantially when we increase the alphabet size from $N$ to $2N$. 
    Thus, in the results that follow in Section~\ref{sec:eval}, we use a latent variable of size $F=N$.

% \vspace{-2mm}
\subsection{Blue-Printing Interference}
\label{sec:blueprinting}
\begin{figure}
    \centering
    \includegraphics[width = \linewidth]{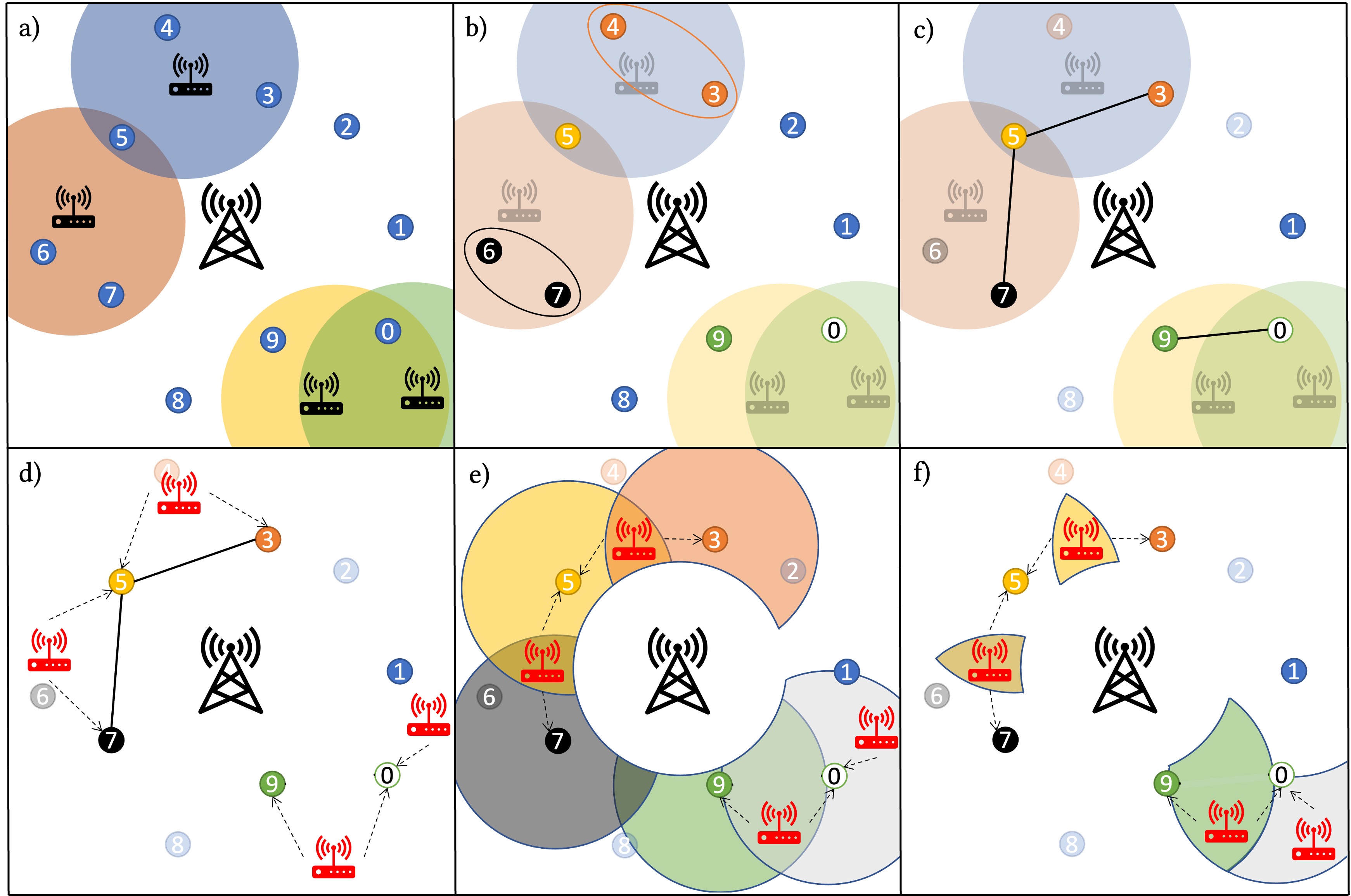}
    \caption{Blue-printing interference and localizing jammers: a) True interference map, 
    b) HOD clustering, 
    c) Graph creation, 
    d) ID HTs, 
    e) Estimated zones, 
    f) Refine zones.}
    % \vspace{-5mm}
    \label{fig:blueprintingandjammer}
\end{figure}

 % \vspace{-1mm}
The estimate of the HOD is sufficient for scheduling as we will see in Section~\ref{sec:resource_mgmt}, but it also provides the basis for generating a more accurate joint description of clients, and consequently interferers, in the physical space. 
Instead of treating the clients as users that are subject to interference, we change our perspective and use them as anchors/sensors with a finite search/impact radius. If we can successfully identify and assign interferers to groups of clients based on their impact, we can create an interference graph. Leveraging the interference graph along with location information of a limited set of clients, we can narrow down and locate HTs to a small, useful area. This has implications for jammer localization, as we will see in Section~\ref{sec:jammer}.
%, and in turn can also bring robustness to resource management in the presence of client mobility. 
%search to an appreciable zone/boundary.
%
We break down the process of generating the map of HTs into two steps, and illustrate using Fig.~\ref{fig:blueprintingandjammer}. 

\textit{\underline{Step 1: HT-focused client clustering:}} 
Define $\textbf{p}_i = \boldsymbol\lambda^* \circ \textbf{P}^*(i,:)$ where $\circ$ is the Hadamard product. The vector $\textbf{p}_i \in [0,1]^F$ is a representation of the joint access probability distribution of client $i$ and the realizations of the latent variable. Joint access vector of two clients with high correlation over the realizations of $H$, implies % which leads us to believe 
that the clients are affected by the same sets of HTs.
On each channel, we form clusters of clients based on these vectors $\textbf{p}_i$ using DBSCAN~\cite{dbscan}.  
In the associated example, this results in the clusters \{0\}, \{1, 2, 8\}, \{3,4\}, \{6,7\}, \{9\}, each colored differently, on the selected channel (Fig.~\ref{fig:blueprintingandjammer}b).
We then identify a representative for each of the clusters at random. The number of clusters obtained is of the order of HTs active on the channel.

\textit{\underline{Step 2: Estimating and mapping HTs: }}
Next, we characterize how groups of clients are jointly impacted. We first generate a dependency graph $G = (V, E)$ where each representative client $i \in V$ and $E = \{(i,j) | \mathbb{P}(i,j) \neq \mathbb{P}(i)\mathbb{P}(j)\}$. That is, an edge exists between two representative clients if the product of their first-order marginals is not equal to the pair-wise marginal, indicating that the channel access probabilities are not independent. A dependency graph is created based on these dependence relationships. In this example, the edges of the graph are \{3,5\}, \{5,7\} and \{0,9\} (Fig.~\ref{fig:blueprintingandjammer}c).

We then take the following approach to approximate the number of HTs active on a channel. We note that 
% \vspace*{-1mm}
\begin{equation}
    \mathbb{P}(\overline{g} | V\backslash g) = f({\small\text{HT impacts at least one of $g$ and not $V\backslash g$}})
    \label{eq:prob_g}
% \vspace{-1mm}
\end{equation}
This equation encodes the relationship between a HT and the likelihood it uses the same channel as the set of clients in $g$. The function $f(\cdot)$ evaluates to $0$ if and only if there is no such HT. This is the key to identifying the smallest descriptive set of HTs.
Using %the relationship established in 
Eq.~\ref{eq:prob_g}, we obtain all such sets of clients affected \textit{jointly} with non-zero probability. We do this by iterating over cliques $g$ generated by graph $G$. We first consider the smallest size cliques and increment the size in each iteration---individuals, pairs, triples, etc.---and add a HT affecting the clique when the following is satisfied:
% \vspace{-2mm}
\begin{equation}
    \mathbb{P}(\overline{g} | V \backslash g) > \prod_i \mathbb{P}(\overline{g_i} | V \backslash g) \hspace{5mm} \forall \text{ tuples } \{g_i\} 
    % \vspace{-2mm}
    \label{eq:prob_compare}
\end{equation}
where $\cup_i g_i = g$ and $g_i$'s are sets found  to be affected by HTs satisfying Eq.~ \ref{eq:prob_g} in a previous iteration. Intuitively, this process identifies whether the clique $g$ is affected by a \textit{new} HT rather than an independent combination of the \textit{existing} ones.  This procedure is summarized in Algorithm~\ref{alg:findHT}.

In the accompanying example, we start with a clique size of 1. The only probability greater than 0 is the probability that client 0 is blocked conditioned on all others transmitting. Hence, we introduce a HT impacting client 0 (Fig.~\ref{fig:blueprintingandjammer}d HT furthest to the right). In the subsequent iteration (clique size 2), we consider the cliques \{3,5\}, \{5,7\} and \{0,9\}.  For each, the probability of being blocked while other clients can transmit is greater than zero. Hence, we introduce three additional HTs to explain the additional information (Fig.~\ref{fig:blueprintingandjammer}d remaining HTs).
\begin{algorithm}
    \caption{Estimating Hidden Terminals}\label{alg:findHT}
    \KwData{Graph G = (V,E), $\boldsymbol\lambda^*, \textbf{P}^*$}
    \KwResult{Set of tuples ($g$) affected by unique hidden terminals}
     Q = formCliques(V,E)\;
     g = emptyList() \Comment*[r]{\small Initializing $g$}
     tuples = formTuples(g) \Comment*[r]{\footnotesize set of all comb. of $g$}
      \For{q \KWIn Q}{
        \For {tuple \KWIn sortBySize(tuples)}{
            \If {q == Union(tuple) }{
                \If {$\mathbb{P}(\overline{q}|V\backslash q) > \prod_i \mathbb{P}(\overline{tuple_i} | V\backslash q)$}{
                    g.append(q); tuples = formTuples(g)\;
                }
            }
        }
      }
\end{algorithm}

The result is to produce a topology like the one in Fig.~\ref{fig:blueprintingandjammer}d which maps four HTs to their respective impacted clients.  %Beyond estimating the CIC, While the locations of the HTs and clients are not known at this point. 
In Section~\ref{sec:jammer}, we show how the estimated interference graph can be leveraged along with a limited amount of knowledge of the clients location to assist in localizing the HTs (shown in Fig.~\ref{fig:blueprintingandjammer}e and f).

%% file: practice.tex
\section{Leveraging \system in Practice}
\label{sec:practice}
We now describe two use cases in which knowledge of interference can significantly benefit the network: resource management and localization of adversarial jammers.
% \vspace{-1mm}
\subsection{Application in Resource Management}
\label{sec:resource_mgmt}
% \vspace{-1mm}
\subsubsection{Current LTE/NR Schedulers} 
% \aadesh{Talk about dynamic traffic}
We briefly capture the essence of existing cellular schedulers where the BS is responsible for allocating time-frequency resource blocks (RBs) to clients on both DL and UL. Scheduling policies that balance throughput and fairness, are typically derived using a utility-optimization approach. We consider the widely-adopted proportional fair (PF) scheduling policy~\cite{kushner_convergence_2004}. 

For SISO systems with $B$ RBs and $N$ single-antenna clients,  the schedule is a matrix $\mathbf{X}\in \{0,1\}^{N\times B}$ that assigns the RBs to users in each time slot. If the set of all possible schedules is $S$, the optimal schedule aims to maximize the aggregate marginal utility:
% \vspace{-2mm}
\begin{equation}
\small
    \mathbf{X}^*(t) = \argmaxA_{\mathbf{X} \in S} \Biggl\{ \sum_{b=1}^{B}  \sum_{i=1}^{N} \frac{x_{i,b} r_{i,b}(t)}{R_i(t-1)} \Biggr\},
    %\text{, s.t. } 
    \sum_{i=1}^{N} x_{i,b} = 1, \forall b
    % \vspace{-1mm}
    \label{eq:propfairSISO}
\end{equation}
where $r_{i,b}(t)$ and $R_i(t-1)$ are the instantaneous rate (on resource block $b$) and average throughput of client $i$, respectively, and 
the constraint on RB assignment $x_{i,b}$, ensures that only one user is scheduled on each RB.

For MU-MIMO systems, where the BS has several antennas ($M$), a group of users is selected, $G \subset \{1, \ldots, N \}$, $|G| < M$,  and  resource blocks are allocated to the group. Here, the optimal schedule
%\footnote{The argument $g$ is necessary because the instantaneous rates $r_i$ is a function of group assignment.} 
is:
% \vspace{-2mm}
\begin{equation}
    \mathbf{Y}^*(t) = \argmaxA_{\mathbf{Y} \in S} \left\{ \sum_{b=1}^{B}  \sum_{i=1}^{N} \frac{y_{i,b} r_{i,b}(t,G)}{R_i(t-1)} \right\}, 
    % \vspace{-1mm}
    \label{eq:propfairMU}
\end{equation}
where $G = \{i|y_{i,b} > 0\}$ and  $\sum_{i=1}^{N} y_{i,b} \leq M, \forall b$.
It is important to note that the MU-MIMO rate of a client $r_{i,b}(t,G)$ varies based on the other clients in its MU-MIMO group. 
At the end of every schedule, the average throughput of all clients is updated through a weighted moving average, to accommodate dynamic traffic conditions. 

\begin{comment}
   
The equations can be solved independently for each resource block. After solving and implementing the schedule, the time averaged throughput of each client $i$ is updated exponentially as
\begin{align}
    R_i (t) \underset{SU-SISO}{   =   } &\alpha \sum_{b=1}^B x^*_{i,b}r_{i,b}(t) + (1-\alpha)R_i(t-1) \label{eq:throughputSISO} \\
    R_i(t) \underset{MU-MIMO}{   =   } &\alpha \sum_{b=1}^B y^*_{i,b}r_{i,b}(t,G) + (1-\alpha)R_i(t-1) \label{eq:throughputMIMO}
\end{align}
where $\alpha \ll 1$ is time-decaying quantity. 
\end{comment}

\subsubsection{Interference-aware Scheduling} 
The existing schedulers, originally designed for licensed spectrum, do not incorporate interference information.  In the following simulations, we progressively add interference information to the scheduling process. We consider the case of uplink, where the impact of HT interference is more pronounced. When each device (including both clients and HTs) employ LBT~\cite{laa_spec} before transmission in unlicensed spectrum, it is possible several of the BS's scheduled UL grants will not be utilized by the clients because of HT interference. 

Recall that we can measure the probability of a client $i$ utilizing its allocated grant, $\mathbb{P}(i)$. 
% , as described in Section~\ref{sec:channelsampling}. 
\begin{comment}
   
The expected marginal utility of schedule $S$ for SISO is then:
\begin{equation}
    E(S) = \sum_{b=1}^B \sum_{i \in S_b} \frac{\mathbb{P}(i) \cdot r_{i,b}}{R_i}.
    \label{eq:expected_marg_util}
\end{equation}

\end{comment}
%
%PF scheduling does not consider the probabilistic nature of interference.  When not accounted for this interference can lead to a significant drop in the resource block utilization in the unlicensed spectrum. One way to compensate for the interference is to rewrite the 
This indirect interference information can be incorporated to benefit the PF scheduler (Eq.~\ref{eq:propfairSISO},~\ref{eq:propfairMU}), 
%Proportional Fair scheduler (Eq.~\ref{eq:propfairSISO},~\ref{eq:propfairMU}) 
so as to maximize the \textit{expected} marginal utility
% function, Eq.~\ref{eq:expected_marg_util}
. 
We term this optimal scheduler \textit{Access-Aware }(AA).  For MU-MIMO the scheduler is:
\begin{equation}
\small
    \mathbf{Y}_{AA}^*(t) \underset{MU-MIMO}{   =   } \argmaxA_{\mathbf{Y} \in S} \Biggl\{ \sum_{b=1}^{B}  \sum_{i=1}^{N} \frac{\mathbb{P}(i) y_{i,b} r_{i,b}(t,G)}{R_i(t-1)} \Biggr\} \label{eq:accessawareMIMO}
\end{equation}
where the same constraints on $y_{i,b}$ from Eq.~\ref{eq:propfairMU} apply. The equation for SISO can be recovered by setting $M=1$.

\subsubsection{Joint Access-Aware Scheduling using HOD for MU-MIMO}
\label{sec:jaa}
%In MU-MIMO systems, an increased number of concurrent transmissions, with sub-optimal precoding, leads to an increased SIR at the receiver \cite{}. \hl{is this right? increased tx w/ bad precoding = improved SIR???}
%On the other hand,
%In an interference rich environment, when a subset of clients cannot transmit because of HTs, the SIR at the BS improves. We model the effect of this increased SIR with a power offset that alters the maximum achievable throughput for a client group depending on the transmissions received.
The aggregate rate of an MU-MIMO group not only depends on the set of users in the group, but also the ability of all users to transmit (subject to interference)---fewer participating streams results in higher SINR for the active streams in the group. 
Therefore, a more accurate model of rate is $r_{i}(t,g,G)$, which is a function of both the group $G$, and the \textit{successful} transmissions $g \subseteq G$. We can compute $\mathbb{P}(g, \overline{G \backslash g})$, i.e. the probability that all clients in $g$ can transmit and the remaining clients, $G \setminus g$ cannot because they are blocked by interference using the parameters $\boldsymbol \lambda, \textbf{P}$ found in Section~\ref{sec:hod_primer}.
%through Eq.~\ref{eq:optimalparameters}.
%We incorporate the new rate model and corresponding probabilities into AA and term 
The resulting scheduler is termed \textit{Joint-Access-Aware} (JAA):
% \vspace{-2mm}
\begin{equation}
    \mathbf{Y}_{JAA}^*(t) = \argmaxA_{ \mathbf{Y} \in S} \Biggl\{ {\small \sum_{b=1}^{B}  \sum_{i=1}^N \sum_{g \subseteq G} \frac{\mathbb{P}(g, \overline{G \backslash g}) y_{i,b} r_{i,b}(t,g,G)}{R_i(t-1)}}  \Biggr\} 
     \label{eq:jointaccessaware}
\end{equation}
Choosing a user-group is non-trivial to solve computationally. We adopt a greedy approach to approximating the solution in a tractable time complexity (refer Eq.~\ref{eq:iterate}). The difference between AA (Eq.~\ref{eq:accessawareMIMO}) and JAA (Eq.~\ref{eq:jointaccessaware}) is that AA assumes that all client transmissions in the group $G$ are successful, which is sub-optimal.

\subsubsection{Speculative Scheduling using HOD for SISO}
\label{sec:sched_sp}
With the JAA  reducing to AA in a SISO scenario, one might question the value for HOD in SISO systems. Note that while AA incorporates access information to cope better with interference, it is not able to efficiently address it and maximize usage of its resource blocks.
%Observe that in a single-user scenario, the \textit{Joint-Access-Aware} scheduler reduces to \textit{Access-Aware}. While the \textit{Access-Aware} scheduler is more optimal than proportional fair, it is not able to maximize the usage of resource blocks. The challenges of interference management highlight the need to deal with scheduling in a more creative way. 
Through a policy termed \textit{Speculative Scheduling} (SP) (introduced in \cite{blu}), we aim to leverage the additional information on these interferers %to make better decisions through 
to adaptively \textit{over-schedule} clients on a RB. Overscheduling is analogous to a popular phenomenon of airlines overbooking flights in expectation that a certain number of people will not make it, leaving the exact number of seats occupied. 

AA limits its vision to single-user scheduling, which naturally restricts the upper bound on RB utilization, determined by interference. We observe that different clients in the same cell can be interfered  by different sets of HTs, leading to {\em interference diversity}. This creates opportunities to schedule multiple clients on the same RB with the expectation that only one of them will be able to transmit, preventing collisions. We primarily consider SISO since the opportunities for over-scheduling diminish with higher multiplexing (MU-MIMO).

For instance, consider a client $i = \argmaxA_{k} \bigl\{\frac{\mathbb{P}(k) r_{k,b}}{R_k} \bigr\}$ that was selected by AA. The speculative scheduler tries to select an additional client $j$ such that scheduling it on the same resource block will likely increase the utility because they are impacted by different HTs. Specifically:
% \vspace{-2mm}
\begin{equation}
    j = \argmaxA_{j \neq i} \biggl\{ \mathbb{P}(i, \overline{j}) \frac{r_{i,b}}{R_i} + \mathbb{P}(\overline{i}, j) \frac{r_{j,b}}{R_{j}} \biggr\}.
    % \vspace{-2mm}
\end{equation}
%If both $i$ and $j$ transmit, it will lead to collision in uplink. 
Note that we want \textit{exactly} one of them to be able to transmit to allow for SISO decoding (lest a collision). For a given set of over-scheduled clients on an RB, its expected utility depends on the joint access probability distribution of the set/group
% \vspace{-2mm}
\begin{equation}
    E(G) = \sum_{i \in G} \mathbb{P}(i, \overline{G \backslash i}) \frac{r_{i,b}(t)}{R_{i}(t-1)}.
    % \vspace{-2mm}
    \label{eq:speculative}
\end{equation}
The optimal scheduler for a SISO system is $\mathbf{G}^*(t) = \argmaxA_{G} E(G)$. 
It is difficult to efficiently compute such groupings, and we resort to a greedy strategy described by Eq.~\ref{eq:iterate}.
We start from a baseline $G$ of two users and updating $G \leftarrow G \cup l^*$ after each iteration.
% \vspace{-2mm}
\begin{align}
    l^* = \argmaxA_{l \notin G} (E(G \cup l) - E(G)) 
    % \vspace{-3mm}
    \label{eq:iterate}
\end{align}

% \vspace{-4mm}
\subsection{Application in Jammer Localization}
\label{sec:jammer}
Our framework for \textit{blue-printing} the network interference from Section~\ref{sec:blueprinting}, % we have formulated a framework to \textit{blue-print} the network interference. Specifically, we have accurately identified 
helped identify the HTs, the sets of clients they impact, and the magnitude of impact. % through Algorithm \ref{alg:findHT}. 
% We refer to this process as \textit{network tomography}. 
An immediate use-case of such HT-client mapping is to identify the \textit{location} of the HTs.   This is of special interest when the HTs are adversarial and the interference is an active, yet subtle (not always ON) attempt to impact the network.

%Using a path-loss model, we can estimate the maximum range between a HT and a client given an energy detection threshold. Given this range (D) and the locations of a few clients (representatives identified in Section~\ref{sec:blueprinting}), we are able to construct an approximate physical map of the interferers. While this indirect localization is not exact, it is nonetheless impressive (median accuracy under 10m, Sec.~\ref{}), and serves our purpose. %is useful for applications in 
%\color{red} list applications \color{black}. 

We accomplish this by changing the anchoring of the interference from clients to physical space. Specifically, we 
%To determine the location of jammers, we need to work with at least the set of representatives, $R$, from each cluster. Next, we 
request the coordinates of representatives $\{(x_i, y_i)\}_{i \in R}$, one each from the clusters identified through Alg.~\ref{alg:findHT}. Now, the region of impact around a particular client $i$ is an area called $A_i \coloneqq \{(x,y)| \text{distance}((x,y),(x_i,y_i)) < D)\}$,
where $D$ is an estimate of the maximum range between a client and a HT, given the client's energy detection threshold (and maximum transmit power in the band). 
A HT located in the area $A_i$ is assumed to impact the access of client $i$, located at $(x_i, y_i)$, while a HT outside it does not. %Further, a HT not affecting the client will not be in this area. 
Now, given a HT obtained using Alg.~\ref{alg:findHT} and the group of clients it impacts ($g$), we can narrow down its location  to the following area,
% \vspace{-2mm}
\begin{equation}
    A_{HT} = \{\cap_{i \in g} A_i\} \bigcap \{\cap_{i \in R\backslash g} A_i^c\} \bigcap A_{BS}^c
    % \vspace{-2mm}
\end{equation}
 where, $A_i^c$ is the complementary area to $A_i$ and $A_{BS}$ is the area around the base station with range $D$. In other words, the HT is restricted to the common area determined by its affected clients and outside the area of both its unaffected clients as well as the BS (by definition of a HT). This process is captured in Fig.~\ref{fig:blueprintingandjammer}e and f. 
 We take the centroid of the resulting area to be the HT's location. Thus, with the locations of a few clients, we are able to construct an approximate physical map of the interferers. While this indirect localization is not exact, it is nonetheless impressive (median accuracy under 10m, Section~\ref{sec:results_jammer}) and sufficient in our application.
 %We also note that a HT must exist outside the detection area of the base station.  The process of localizing jammers is shown in Fig.~\ref{fig:blueprintingandjammer}e and f.

%% file: evaluation.tex
\section{Evaluation}
\label{sec:eval}
\vspace{-1mm}
We evaluate \system comprehensively in both resource management and jammer localization applications using a combination of NS3~\cite{ns3}  and numerical simulations. %While we are able to instrument NS3 for DSA on a single channel SISO system using the LAA-LBT implementation \cite{ns3-laa}, it is not conducive for multi-channel and multi-antenna scenarios, wherein we resort to numerical simulations. 

%\karthik{Do we indicate what LAA-LBT is anywhere before? I included a reference to LBT in Sec 2.}
%\aadesh{Yes, we do in section 5.1.2. But I added a reference in motivation}
We are able to instrument NS3 for DSA on a single channel SISO system using the LAA-LBT implementation \cite{ns3-laa}. In these simulations, we consider 100 topologies (similar to Fig.~\ref{fig:HT-channel}). The HTs are distributed uniformly, but to ensure they are ``hidden" from the BS, they are placed at least $70 m$ away, where $50 m$ is approximately the radius of the HT's zone of impact. There are 2 to 8 HTs (Wi-Fi APs) randomly distributed amongst 8 to 20 clients. The BS is centrally located and the Wi-Fi nodes are out of its range, but impact the clients. NS3 currently supports LBT only at the BS (not at clients). Hence, we incorporate the impact of collisions using the physical layer traces. 
%does not yet support LThe LBT implementation we work with does not have an LBT implementation on the clients, so we adjust for collisions using the physical traces. 
While this works for SISO, it is not conducive for multi-channel and MU-MIMO scenarios with many clients, wherein we supplement with numerical simulations. Our topologies for numerical evaluations mimic our NS3 network set-up. 
%
% \karthik{why only distance info for numerical, and not for NS3, if they are same, mention it for NS3 and shorten description for numerical}
In all our simulations, clients are uniformly distributed in a circular area bounded by the BS's transmission range.  
% estimating

To estimate the HOD, the BS follows the procedure in Section~\ref{sec:system_design}.  The first- and second- order marginals are estimated over 1000~frames.  Unless specified, the number of clusters is $N/C$, for $N$ clients and $C$ channels.
% p HT
Each HT has a fixed probability of transmission drawn uniformly from the interval $[0.2, 0.8]$.  If any node, either a client or another HT, is within the HT's zone of impact, the probability that they will not be able to transmit at a given frame is dependent on the HT transmit probability.
% scheduling
Simulations consists of scheduling 1,500 LTE/5G frames.  In each frame the BS attempts to schedule the $N$ clients on $B=10$ available RBs according to the schedulers presented in Section~\ref{sec:resource_mgmt}.

% \subsection{NS3 Implementation}
\label{sec:ns3}
%We instrument the LTE-LBT implementation \cite{ns3-laa} in NS3 to test \system's performance in resource allocation as well as jammer localization applications.

\begin{figure}
    \centering
     \begin{subfigure}[b]{0.39\linewidth}
         \centering
        \includegraphics[width=\textwidth]{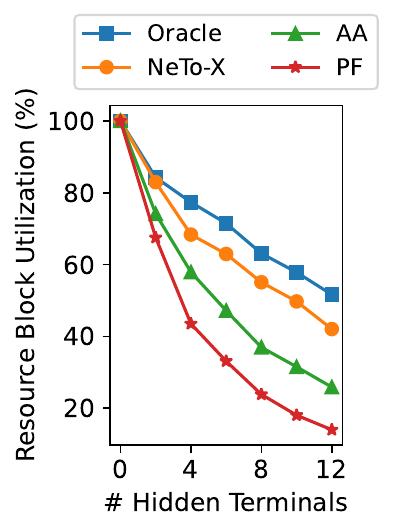}
         % \caption{$y=3\sin x$}
     \end{subfigure}
    \begin{subfigure}[b]{0.59\linewidth}
         \centering
        \includegraphics[width=\textwidth]{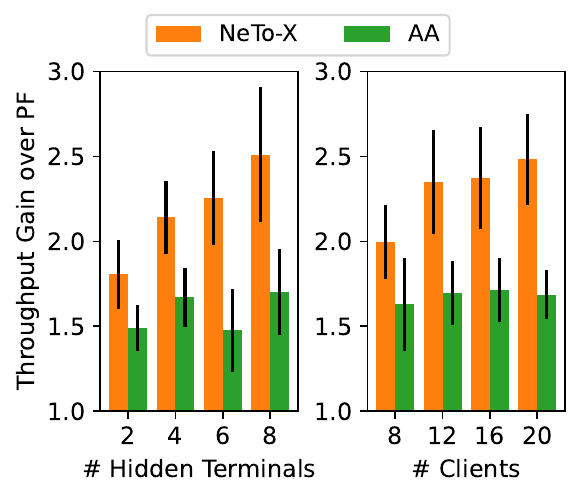}
         % \caption{$y=x$}
     \end{subfigure}
     \caption{a) RB utilization on a single channel SISO system, b) Throughput gains over PF with varying HTs and c) clients.}
     % \vspace{-5mm}
     \label{fig:ns3}
\end{figure}

% \vspace{-2mm}
\subsection{Resource Management}
\label{sec:results_sched}
% \vspace{-1mm}
First, we look at how the estimated HOD can be used to optimize client scheduling.
We compare \system to PF and AA schedulers as described in Section~\ref{sec:resource_mgmt} as well as an optimal \textit{oracle} scheduler. Our metrics include resource block utilization (RBU), which captures how efficient the scheduler is.  We also report a measure of the cumulative throughput a schedule can achieve which is simply the sum of the long-term rates, and is normalized to that achieved by PF. Note that fairness is automatically incorporated by the schedulers in their policy. Additionally, we investigate the trade-off between performance and the measurement overhead.

% We vary the number of clients in Fig.~\ref{fig:ns3}b while keeping the number of HTs fixed at 3. When the number of clients increases, \system identifies more groups of clients that could be scheduled together which results in more opportunities to overschedule and thus greater throughput gain over PF. 
\subsubsection{Impact of Interference}
\label{sec:results_sched_ht}
% \begin{figure}
%     \centering
%     \includegraphics[width = 0.45\linewidth]{plots/plot-HT-small.pdf}
%     \caption{Numerical simulations with a 4 channel system}
%     \label{fig:HT}
%     \vspace{-0.2in}
% \end{figure}
We aim to understand the scheduler performance with increasing levels of interference.
Fig.~\ref{fig:ns3}a and b capture the effect of adding hidden terminals on the same channel for 20 clients using NS3. 
Fig.~\ref{fig:ns3}a shows the percentage of resource blocks successfully utilized in a single-channel SISO system for 20 clients.
Notice that the resource block utilization of all schedulers, including the \textit{oracle} decrease as the level of interference increases, highlighting the inherent challenges of DSA.
The PF scheduler performs worse since it is oblivious to the interference. 
Beyond eight HTs, its efficiency falls below 25\%.  The AA scheduler, which simply weights the PF schedule with the measured first-order marginal channel access probabilities, performs only slightly better.  
However, by estimating the HOD with \system, and performing speculative scheduling, we are almost able to achieve the the performance of the \textit{oracle}, differing by only a few percent.
In Fig.~\ref{fig:ns3}b, with an increase in the number of hidden terminals, 
the cumulative throughput falls drastically for PF (from (a)), 
\system minimizes losses through speculative scheduling to yield an appreciable gain of 50\% over AA and 2x over PF.

% QUESTION: The graph looks like it could be asymptotic.  Is it?  Is there some kind of hardening to the interference? That could be interesting to observe.

\subsubsection{Impact of Client Density} The LAA-LBT simulations in NS3 support a limited number of clients. In Fig.~\ref{fig:ns3}c, we observe that the throughput gains over PF increase substantially with an increasing number of clients, wherein the diversity of interference impact is more. \system is able to leverage this diversity to speculatively schedule more groups of clients together. These gains are also observed for a much larger client density in our numerical evaluations. 

\subsubsection{Multi-Channel Performance} 
\label{sec:results_sched_chan}
\begin{figure}
    \centering
    \includegraphics [width=0.9\linewidth] {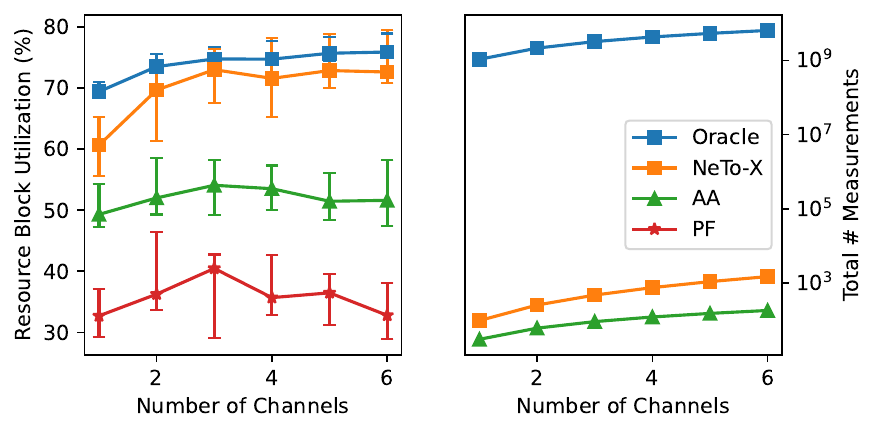}
    % \vspace{-3mm}
    \caption{Resource block utilization per channel and measurement overhead of SISO schedulers.}
    % \vspace{-5mm}
    \label{fig:channel}
\end{figure}
Next, we would like to investigate how the number of channels impact scheduling performance.  Recall that modern LTE/5G networks use multiple channels thorugh carrier aggregation to increase overall throughput, but this increases resource management complexity especially in the interference-rich environment of DSA.

The performance and measurement overhead of a SISO system with 30 clients vs. number of channels is shown in Fig.~\ref{fig:channel}.  To isolate the effect of the channels from the level of interference, the number of hidden terminals per channel is fixed at 4. 
%the number of HTs is increased proportionally to the number of channels (4 HT/channel).
The performance of \textit{PF} and \textit{AA} is independent of the number of channels as they don't utilize cross-channel information. In contrast, \system clusters clients affected together, enhancing joint access probability estimation across channels for more efficient channel allocation. This allows \system to approach the performance of the \textit{oracle} scheduler with fewer measurements (a factor of $10^{-6}$) as the number of channels increases.

\subsubsection{Multi-Antenna Systems}
\label{sec:results_sched_mimo}
\begin{figure}
    \centering
    \vspace{-5mm}
    \includegraphics[width= 0.9\linewidth]{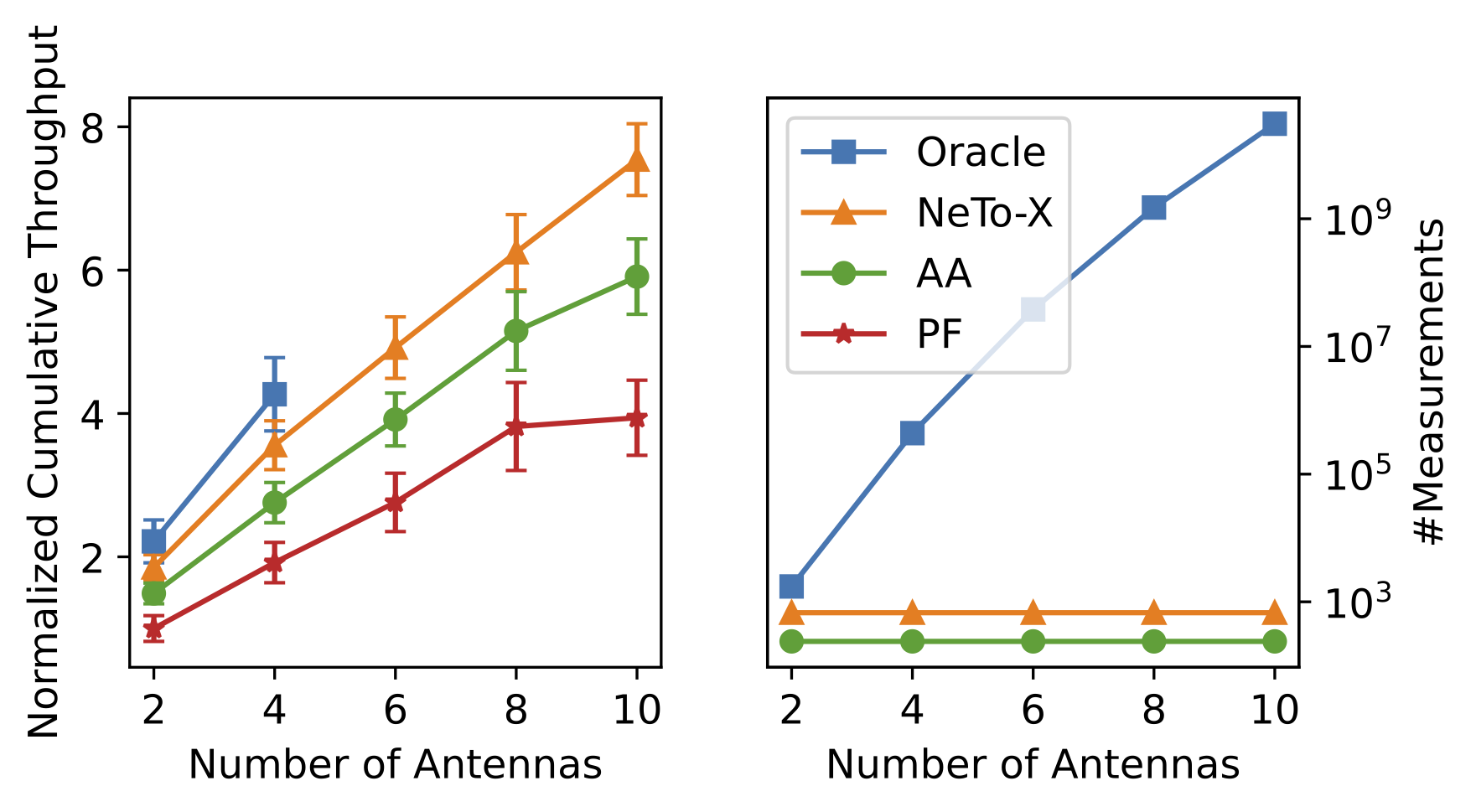}
    \caption{Normalized total throughput and overhead of MU-MIMO vs. number of antennas.}
    % \vspace{-5mm}
    \label{fig:MIMO}
\end{figure}
We now investigate the benefits of \system for scheduling in MU-MIMO systems.
\system outperforms both PF and AA schemes for client scheduling in MU-MIMO systems as seen in Fig.~\ref{fig:MIMO}.  
Indeed, the cumulative throughput of \system grows linearly with the number of antennas, while the PF scheduler starts exhibiting saturation with more antennas and is unable to deliver the  promised gains.
For a limited number of antennas we are able to compare \system to the \textit{oracle}; however, as the number of antennas increases, the computational and storage overhead, let alone the required number of measurements, increases to a point where even simulation becomes infeasible. 
On the other hand, \system has a significantly smaller measurement overhead which \textit{does not} scale with the number of antennas and clients making its implementation computationally feasible. 

\subsubsection{Clustering Analysis}
\label{sec:cluster_analysis}
\begin{figure}
    \centering
    \vspace{-7mm}    
    \includegraphics[width=0.9\linewidth]{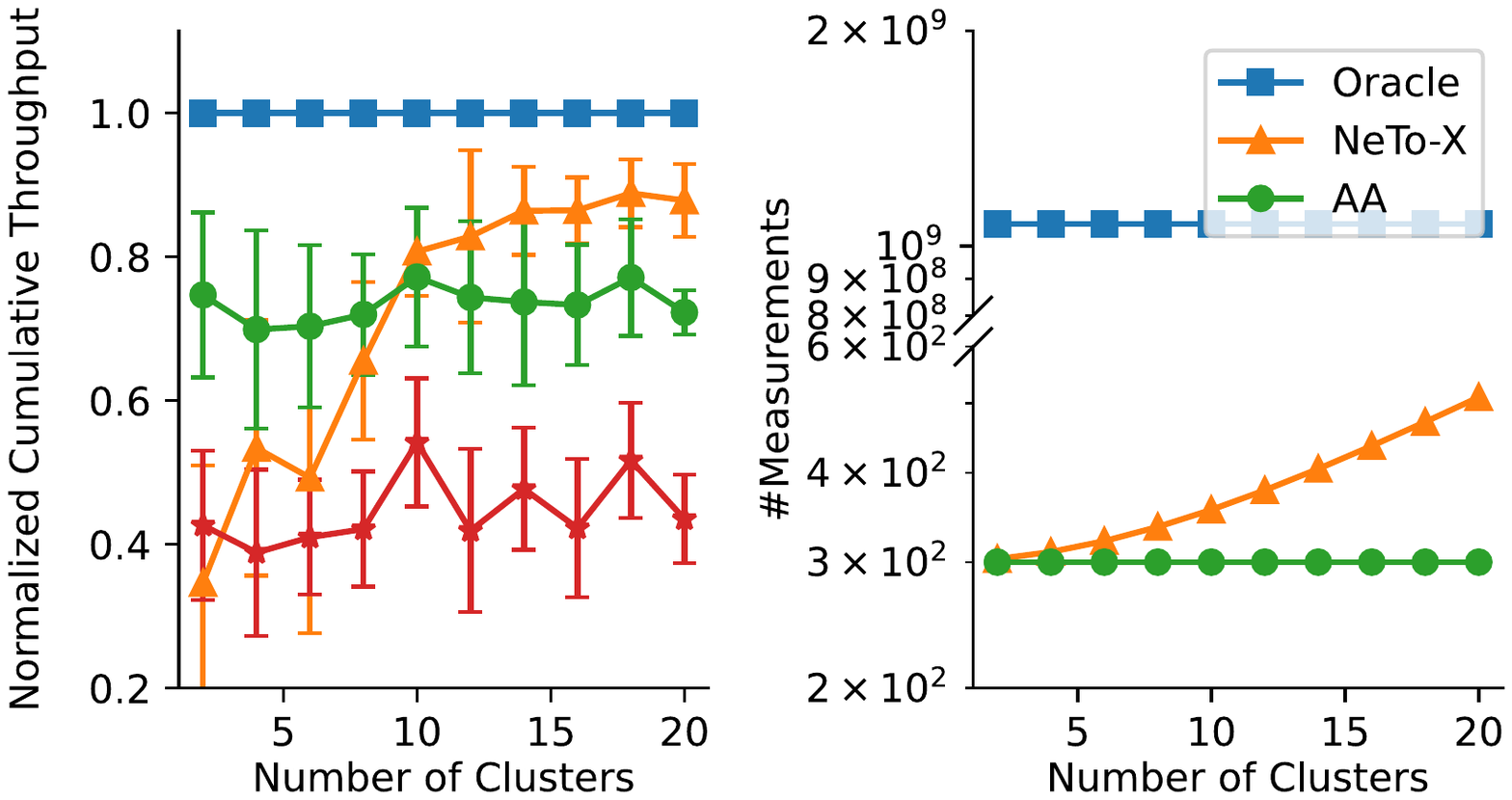}
    \vspace{-7mm}
    \caption{Effect of the number of clusters used by \system on the performance and measurement overhead.}%, $N=150$, $C=5$, $HT/C=4$}
    % \vspace{-5mm}
    \label{fig:clusters}
\end{figure}
One of the design parameters is the number of clusters, $K$, formed after obtaining the first-degree marginals as described in Section~\ref{sec:system_design_overview}.
By grouping clients into clusters, we are making the assumption that the they are impacted by the same HTs.
A smaller number of clusters requires fewer measurements to estimate the pair-wise marginals, but it may mean that we have made incorrect assumptions about the nature of the interference.
The trade off between scheduling performance and measurement overhead as a function of the number of clusters is captured in Fig.~\ref{fig:clusters}.
Aggressive clustering ($K<10$ for 150 clients) results in the least amount of measurements but also causes \system to underperform compared to AA and potentially even PF.
Increasing the number of clusters provides a better estimation of the joint access distribution at the cost of an increase in the number of measurements which scales as $\binom{K}{2}$, though this overhead is still significantly less than that of the \textit{oracle}.

% \vspace{-3mm}
\subsection{Evaluation for Jammer Localization}
\label{sec:results_jammer}
% \vspace{-1mm}
% \karthik{It would benefit to have a small table on median loc error with increasing HTs [2-8]}
To evaluate how well \system can be used to locate jammers, we simulate the presence of 2-8 HTs with 40 clients over 200 randomized topologies. For each topology, we try to localize jammers using the method described in Section~\ref{sec:practice}. We perform the same analysis using the locations of (1) all UEs and (2) only the representatives described in Section~\ref{sec:blueprinting}. The number of representatives ranges from 3 to 10 clients depending on the topology. After narrowing down the HTs to a candidate zone, we measure the metrics
\begin{itemize}
    \item \textit{Accuracy: } the distance between the ground truth location of the HT and the centroid of the candidate zone,
    \item \textit{Precision: } the total area of the candidate zone. 
\end{itemize}

\begin{figure}
    \centering
    \includegraphics[width= \linewidth]{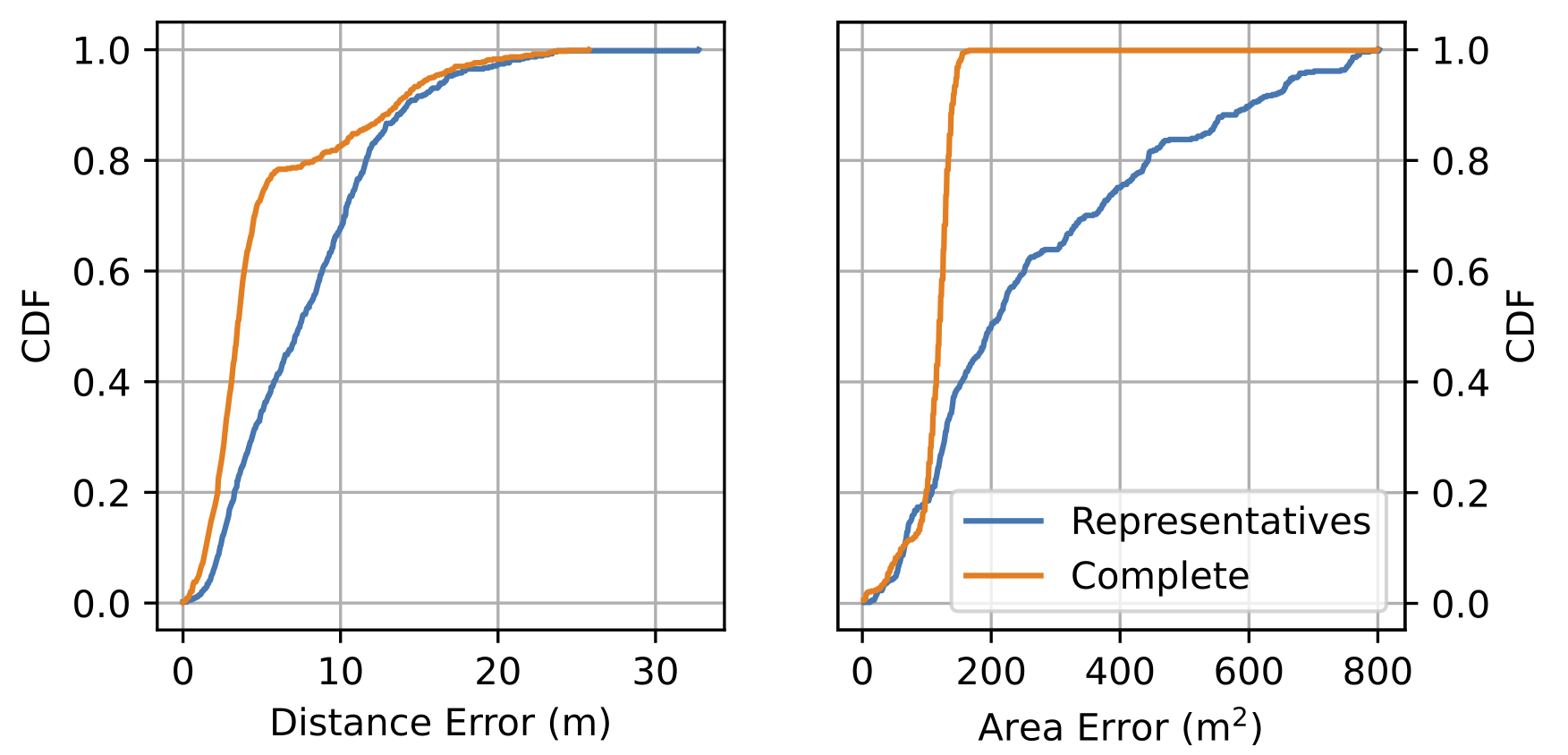}
    \caption{Accuracy (left) and Precision (right) of \system in localizing jammers with 40 clients and 3 hidden terminals}
    % \vspace{-3mm}
    \label{fig:jammer}
\end{figure}

\renewcommand{\arraystretch}{1.15}
\begin{table}[t]
    \centering
    \begin{tabular}{|c||c|c|c|c|c|}
        \hline
        \textbf{Number of HTs}     & 2 & 3 & 4 & 6 & 8 \\
        \hline
        \textbf{Accuracy (\%)}   & 99.3 & 90.5  & 81.7 & 73.5 & 70.6 \\
        \hline
    \end{tabular}
    \caption{Classification accuracy of \system}
    % \vspace{-6mm}
    \label{tab:classification}
\end{table}

Fig.~\ref{fig:jammer} depicts the accuracy and precision. When using all clients' location as anchors, the median accuracy is about 5~m.  Using only the representatives' locations, the median accuracy reduces to 10~m. This granularity is still sufficient to infer the location of the jammers with high likelihood.

However, using the locations of only the representatives makes pinpointing the location of jammers precisely more challenging. The client placement is sparse and narrowing down the candidate zone is difficult. While the median precision is 200 $m^2$, there is a long tail which extends beyond 400 $m^2$ in the 80$^{th}$ percentile. When we have complete information of the locations, \system is able to determine an appreciably smaller candidate zone. 

Table~\ref{tab:classification} shows the accuracy with which \system identifies the number of HTs present on a channel over 200 sample topologies with 40 clients. \system accurately identifies the number of jammers on the network ($>90\%$ for up to 3 HTs). However, for a larger number of HTs, the accuracy dips to 70\% as some HTs are very closely located to each other and tend to present themselves as a single, consolidated HT. However, this might still be satisfactory to localize the area of aggregate HT presence.

%% file: conclusions.tex
\section{Conclusions}
% \vspace{-1mm}
As future wireless networks explore flexible spectrum sharing models, understanding the impact of external interference has become central to delivering the increased spectral efficiencies offered by their multi-antenna, multi-channel solutions.
To this end, we proposed a novel, scalable network tomography
framework called \system that transforms clients into sensors, and leverages their joint channel access statistics to accurately characterize the interference sources and their impact at a highly scalable overhead. Its merits are showcased in the context of
a resource management and jammer localization application, where
its performance significantly outperforms baseline approaches, and
closely approximates optimal performance at significantly (several orders of magnitude) lower overhead.